\documentclass[%
reprint,
superscriptaddress,
amsmath,amssymb,  aps, prc, nofootinbib
]{revtex4-1}
\usepackage{mathtools}
\usepackage{graphicx}% Include figure files
\usepackage{dcolumn}% Align table columns on decimal point
\usepackage{bm}% bold math
\usepackage{subfigure} 
\usepackage{xcolor}
\usepackage{etoolbox}
\usepackage{hyperref}% add hypertext capabilities
\makeatletter
\appto\abstract{%
  \let\latexlist\list
  \def\list{\edef\keeprightskip{\the\rightskip}\latexlist}%
  \patchcmd\latexlist{\ignorespaces}{\rightskip\keeprightskip\ignorespaces}{}{}%
}
\makeatother

\begin{document}

\title{Collision term dependence of the hadronic shear viscosity and diffusion coefficients}

\author{Jan Hammelmann}
\affiliation{Frankfurt Institute for Advanced Studies, Ruth-Moufang-Strasse 1, 60438 Frankfurt am Main, Germany}
\affiliation{Institute for Theoretical Physics, Goethe University, Max-von-Laue-Strasse 1, 60438 Frankfurt am Main, Germany}

\author{Jan Staudenmaier}
\thanks{Has left academia}
\affiliation{Institute for Theoretical Physics, Goethe University, Max-von-Laue-Strasse 1, 60438 Frankfurt am Main, Germany}

\author{Hannah Elfner}
\affiliation{GSI Helmholtzzentrum f\"ur Schwerionenforschung, Planckstr. 1, 64291 Darmstadt, Germany}
\affiliation{Institute for Theoretical Physics, Goethe University, Max-von-Laue-Strasse 1, 60438 Frankfurt am Main, Germany}
\affiliation{Frankfurt Institute for Advanced Studies, Ruth-Moufang-Strasse 1, 60438 Frankfurt am Main, Germany}
\affiliation{Helmholtz Research Academy Hesse for FAIR (HFHF), GSI Helmholtz Center, Campus Frankfurt, Max-von-Laue-Straße 12, 60438 Frankfurt am Main, Germany}

\date{\today}

\begin{abstract}
\begin{description}
\item[Background] The value of the shear viscosity $\eta$ and the diffusion coefficients of conserved charges $\kappa_{ij}$ with $i,j\in\{B,Q,S\}$ strongly depend on the microscopical interactions of the constituents.
\item[Purpose] Study the effect of multi-particle reactions, angular distributions and additional elastic cross sections via the additive quark model description, the numbers of degrees of freedom and the baryon density on the transport coefficients.
\item[Method] $\eta$ and $\kappa_{ij}$ are extracted from infinite matter simulations of the hadronic transport model SMASH using the Green-Kubo formalism. The transport coefficients are then obtained by integrating the correlation functions which has been thoroughly checked and compared to kinetic theory calculations. We then systematically study the collision term dependence of $\eta$ and $\kappa_{ij}$ by switching the respective cross sections on and off. With this method we are able to extract the relevant interactions and dependencies for $\eta$ and $\kappa_{ij}$.
\item[Results] We find that multi-particle reactions decrease the shear viscosity in a simplified hadron gas whereas the electric charge diffusion coefficient is not modified. Furthermore, additional elastic cross sections have a strong impact on both $\eta$ and $\kappa_{ij}$ whereas anisotropic scatterings enhance the shear viscosity in the full hadron gas. When increasing the number of degrees of freedom the shear viscosity is only slightly modified in comparison to the diffusion coefficients. Finally, the calculation within a finite baryon chemical potential reveals that the shear viscosity itself does not depend on $\mu_B$ but on the ratio $\eta / s$. The diffusion coefficients show a strong dependency which we compare to Chapman-Enskog calculations.
\item[Conclusions] We conclude that the behavior of the shear viscosity is highly sensitive to the individual treatment of the cross sections. The diffusion coefficients of conserved charges are affected by the averaged cross sections and the charge densities. These results help understand observables when modeling heavy-ion collisions as well as open the window for a better understanding when comparing those to first principle calculations.
\end{description}

\end{abstract}

\maketitle
\section{\label{intro}Introduction}
  Transport coefficients play an important role in understanding the properties of matter. They describe the reaction of the system to perturbations out of equilibrium back to the equilibrium state and they are governed by the microscopic interactions of its constituents. In the case of quantum chromo dynamics (QCD), there is a large interest specifically in the value of the shear viscosity $\eta$ and the diffusion coefficients $\kappa$ of conserved charges, since they govern the evolution of the hot and dense medium created in relativistic heavy-ion collisions performed at various different collision energies e.g. at the Large Hadron Collider (LHC) or the Relativistic Heavy-Ion Collider (RHIC).
  
  The shear viscosity describes the behavior of matter as a reaction to forces applied on one side of the system. If $\eta$ is very small the matter is called perfect or ideal, meaning that force gradient are perfectly propagated through the medium.
  Similar to e.g. Helium or $\mathrm{H_2O}$ which has a minimal value of $\eta / s$ around the phase transition \cite{Csernai:2006zz}, it is thought that the minimal value of $\eta / s$ of QCD matter is also in temperature regions of the phase transition between hadronic and partonic degrees of freedom.
  A direct measurement of the shear viscosity of QCD matter is not possible however there are observables in relativistic heavy-ion collision experiments which have been related to $\eta$ such as the anisotropic flow coefficient $v_2$ \cite{Shen:2020gef}. Using hydrodynamical simulations which need the transport coefficients as an input it has been found that the shear viscosity over entropy density $\eta / s$ of the QGP needs to be very small in order to explain experimental results at RHIC or LHC energies \cite{Petersen:2008dd, Gale:2012rq, Petersen:2014yqa, JETSCAPE:2020shq, Auvinen:2017fjw} as well as in lower collision energies \cite{Karpenko:2015xea, Gotz:2022naz, Auvinen:2017fjw}. 
  
  Unfortunately calculations of the shear visosity from first principle calculations are not present at the moment therefore one has to rely on effective model calculations or other methods.
  A well known calculation using the AdS/CFT correspondence found that for many strongly interacting quantum field theories there exist a lower bound of the shear viscosity over entropy density ratio $\eta / s$ of $1 / 4\pi$ \cite{Kovtun:2004de} (KSS bound).
  
  There are multiple studies calculating the value of the shear viscosity in the hadronic regime of QCD using transport models \cite{Muronga:2003tb, Demir:2008tr, Wesp:2011yy, Plumari:2012ep, Ozvenchuk:2012kh, Pratt:2016elw}, chiral perturbation theory \cite{Torres-Rincon:2012sda} or other methods like HRG with excluded volume effects or \cite{McLaughlin:2021dph}.
  In \cite{Rose:2017bjz} the effect of lifetimes of resonances on the shear viscosity was discussed and its importance when comparing results between different models. This result shows that the treatment of microscopic interactions plays an important role when calculating the shear viscosity.

  There is a large effort to use Bayesian analysis techniques to extract the shear viscosity over entropy density from experimental data \cite{Bernhard:2016tnd, JETSCAPE:2020mzn, JETSCAPE:2020shq, Nijs:2020roc, Parkkila:2021tqq, Parkkila:2021yha}. The value of $\eta / s$ has been found to be close to the KSS bound. Other methods to extract the shear viscosity from dynamical models shows a similar picture \cite{Reichert:2020oes, Yang:2022yxa}.

  A different class of transport coefficient which receives more attention in recent years are the diffusion coefficients of the baryonic $B$, electric $Q$ charge and strangeness $S$ as well as their cross coefficients $\kappa_{ij}\; i,j\in\{B,Q,S\}$. In general, diffusion coefficients determine the ability of the medium to disperse charges throughout the system, if there are inhomogeneities in the initial state of some evolution.
  Since many hadrons carry more than one charge e.g. a proton has both baryon and electric charge, a baryon current will therefore create an electric charge current and vice versa. The strength of their coupling can be expressed by the cross diffusion coefficients. Coming back to heavy-ion collisions, charge diffusion processes are expected to have a large effect at lower collision energies in contrast to ultra relativistic energies where the net baryon density is approximately zero around midrapidity \cite{Denicol:2018wdp, Fotakis:2019nbq}. 
  It has been shown that at LHC energies the baryon diffusion only has a effect in the longitudinal direction \cite{Monnai:2012jc}.

  The first calculation of the full cross diffusion coefficient matrix $\kappa_{ij}$ including the cross terms of a hadron gas was performed in \cite{Greif:2017byw} within an approach to kinetic theory.
  A similar calculation within the Chapman-Enskog (CE) approximation and the relaxation time approximation (RTA) for the transition region between hadronic and partonic sector of QCD was performed in \cite{Fotakis:2021diq}.
  Contrary to the shear viscosity $\eta$, the electric conductivity $\sigma_{el}$ which is equal to $\kappa_{QQ} / T$ as well as the diffusion coefficient $D$ have been calculated from first principle lattice QCD calculations \cite{Aarts:2014nba, Ding:2016hua}.
  Within the transport model that is also used in this work, a calculation of the electric charge conductivity and respective cross-terms ${QQ, QB, QS}$ was performed in \cite{Rose:2020sjv}.
  Further studies that also used transport models of the electric conductivity can be found in \cite{Puglisi:2014sha, Greif:2014oia, Hammelmann:2018ath}.

  A systematic study and the phenomenological implications of the inclusion of baryon diffusion in hydrodynamical modeling of heavy-ion collisions at intermediate beam energies was done in \cite{Denicol:2018wdp}. 
  In \cite{Fotakis:2019nbq} a hydrodynamical calculation was performed which incorporated the full diffusion coefficient matrix and it was observed that the net-strangeness density in the longitudinal direction depends on the full matrix $\kappa_{ij}$. It has been found that locally, regions of non-zero net strangeness density evolve during the hydrodynamical evolution.
  
  In this work we want to extend previous studies and focus on the collision term dependence of the transport coefficients. Specifically, the impact of multi-particle reactions as well as angular distributions and the incorporation of additional cross sections via the description of the additional quark model (AQM) are studied. We also investigate the effect of varying hadronic degrees of freedom on the shear viscosity and diffusion coefficients.
  Recently the model we use was extended such that specific hadronic multi-particle reactions can be studied which enables us to investigate the effect of e.g. $3\leftrightarrow 1$ processes on transport coefficients for the first time in the hadronic medium. This collision criterion has been used in a partonic transport model for the description of gluon bremstrahlung reactions \cite{Xu:2004mz, Xu:2007aa}.
  Additionally, the influence of angular distributions on the shear viscosity has not been studied in the hadronic stage yet whereas in the partonic medium it has been pointed out to be important as it increases the shear viscosity over entropy density ratio \cite{Lin:2021mdn}.
  In \cite{Puglisi:2014sha} it was found that strongly anisotropic distributions of outgoing particles increases the electric conductivity.
  Finally the effect of additional cross sections will be investigated. They have been recently introduced into the model and are assumed to be important as they estimate cross sections between particles where there exist no experimental measurements. 

  The rest of this work is organized as follows. First the model is introduced which is used to extract the shear viscosity and diffusion coefficients.
  In Sec. \ref{methodology} the Green-Kubo methodology is described. In Sec. \ref{Fig:EtaMultiPart} the impact of multi-particle reactions on $\eta$ and the electric charge diffusion coefficient $\kappa_{QQ}$ is calculated and in Sec. \ref{Fig:EtaOverSFullSMASH} the influence of angular distributions and AQM cross sections on $\eta$ and $\kappa_{ij}$ are discussed. We continue with a calculation on the impact of the degrees of freedom in the model in Sec. \ref{Sec:DOF} and finally move to results of having a non-zero baryon chemical $\mu_B$. A summary of this work is presented in Sec. \ref{Sec:FiniteMuB}.

\section{\label{smash}SMASH transport approach}

To extract the equilibrium value of the shear viscosity of hadronic matter as a function of temperature, we employ the hadronic transport approach SMASH \cite{Weil:2016zrk,dmytro_oliinychenko_2020_4336358,SMASH_github}. SMASH is build to simulate heavy-ion collisions at intermediate collision energies and has been successfully used to extract various transport coefficients from conductivities to bulk and shear viscosities \cite{Rose:2017bjz,Hammelmann:2018ath,Rose:2020sjv,Rose:2020lfc,Dorau:2019ozd}. The extraction of transport coefficients is performed in a box with periodic boundary conditions simulating infinite matter. The hadrons are described by point like particles and their interactions with each other are governed by cross sections of specific interaction channels.

An important ingredient of transport approaches is the collision criterion, which decides when interactions happen. A common criterion is the geometric collision criterion 
  \begin{equation}
    d_{\rm{int}} = \sqrt{\frac{\sigma_{\rm{tot}}}{\pi}} \, .
  \end{equation}
Here, two particles interact when their transverse distance is smaller than $d_{\rm{int}}$, which is given by the geometric interpretation of their total cross section. In SMASH a covariant form of the geometric criterion was introduced in the code \cite{Hirano:2012yy}.
If the criterion is fulfilled, the specific interactions e.g. elastic or inelastic scattering, is chosen depending on the partial cross section of the subprocess.
The geometric criterion is limited to two interaction partners as the generalization of the interaction time and transverse distance to more particles is difficult.
It is however theorized that in a dense medium reactions with more than two reaction partners i.e. multi-particle interactions become a relevant part of the dynamics. In addition, there are well known hadronic interactions that exceed two particles in the initial or final state e.g. $\omega\leftrightarrow 3\pi$ or $N\bar N \leftrightarrow X\pi$. Even though the decay and the $N\bar N$ two-body reaction could be accounted for with a geometric criterion, the 3- or 5-body reaction would need to be neglected, breaking detailed balance. The only option is to model such multi-particle interactions with binary interactions introducing intermediate resonance states into an reaction chain. For example using the $\rho$ meson in $\omega\leftrightarrow\pi\rho\leftrightarrow 3\pi$.

In order to treat multi-particle reactions directly, it is necessary to use a different collision criterion that is easily generalized to interactions with more than two incoming particles, the stochastic collision criterion \cite{Cassing:2001ds, Seifert:2017oyb, Xu:2004mz}. It was was recently introduced in SMASH~\cite{Staudenmaier:2021lrg,Garcia-Montero:2021haa}.
 
For the stochastic collision criterion, a collision probability is defined for each possible reaction within a phase space cell $\Delta^3x \Delta^3p$ within a given timestep $\Delta t$. This  probability can be directly introduced for a $n\rightarrow m$ reaction as
  \begin{equation}
    P_{n\rightarrow m} = \frac{\Delta N_{\rm{reactions}}^{n\rightarrow m}}{\prod_{j=1}^n \Delta N_j} \, .
  \end{equation}
  Here, $\Delta N_j$ is the number of particles within the cell and $\Delta N_{\rm{reactions}}^{n\rightarrow m}$ the number of reactions for the timestep within the cell. With the scattering rate given by the collision term of the Boltzmann equation, the probability for 2- and 3-body reactions can be expressed in terms of cross section or decay width of the reverse process. More details of this derivation and the numerical treatment are found in \cite{Staudenmaier:2021lrg}. 
  
The probability of a arbitrary 2-to-m scattering is given by 
\begin{equation}
P_{2 \rightarrow m} = \frac{\Delta t}{\Delta^3 x} v_{\textrm{rel}} \sigma_{2\rightarrow m} (\sqrt{s})     
\end{equation}
with the timestep size $\Delta t$, the cell volume $\Delta^3x$, the cross section of the process and the relative velocity
\begin{equation}
v_{\textrm{rel}} = \frac{\lambda^{1/2}(s;m_1^2,m_2^2)}{2E_1 E_2} \ , \label{eq:vrel}
\end{equation}
which uses the abbreviation $\lambda(s;m^2_1,m^2_2)=(s-m_1^2-m_2^2)^2-4m^2_1m_2^2$.

The probability for a 3-to-1 reaction is given by 
\begin{equation}
P_{3\rightarrow 1}  = \left( \frac{g_{1'}}{g_1 g_2 g_3} \right) {\cal S}! \frac{\Delta t}{(\Delta^3 x)^2} \frac{\pi}{4E_1E_2E_3} 
\frac{\Gamma_{1\rightarrow 3} (\sqrt{s})}{\
\Phi_3(s)} {\cal A} (\sqrt{s})  \ , \label{eq:P31}
\end{equation}
in terms of the decay width of the reverse process $\Gamma_{1\rightarrow 3} (\sqrt{s})$, the spectral function of the formed resonance ${\cal A} (\sqrt{s}) $, the 3-body phase space $\Phi_3(s)$, the spin degenercy factor $g_j=2\mathfrak{s}_j+1$ ($\mathfrak{s}_j$ being the spin of the state) and ${\cal S}$ the number of identical incoming particles.
With this the inverse reactions to mesonic Dalitz decays i.e. 3-to-1 reactions are acoounted for in SMASH in order obtain detailed balance. The  3-to-1 reactions in SMASH are $\pi\pi\pi \rightarrow\omega$, $\pi\pi\pi\rightarrow\phi$ and $\pi\pi\eta\rightarrow\eta'$. Other multi-particle reactions realized in SMASH are the light nuclei generation with 3-to-2 interactions and the back-reaction of $N\bar N$ anihilations via 5-to-2 reactions. Results with them are discussed in~\cite{Staudenmaier:2021lrg} and \cite{Garcia-Montero:2021haa} respectively.

 It is well known that the value of the shear viscosity is strongly dependent on interactions and cross sections and therefore the transport coefficient changes whenever new interactions are incorporated into the model. The two main differences in comparison to \cite{Rose:2017bjz} are first the incorporation of the additive quark model (AQM) \cite{Goulianos:1982vk} cross sections. And second the impact of angular distributions of the final state in binary interactions.
 AQM cross sections give a parametrization for cross section between hadrons which are not known experimentally. They can be written as
 \begin{equation}\label{Eq:AQM}
   \sigma^{AQM}_x = 40 \left( \frac{2}{3} \right)^{n_{\rm{meson}}} (1 - 0.4 x^s_1) (1 - 0.4 x^s_2) \, ,   
 \end{equation}
 where $n_{\rm{meson}}$ is the number of mesons and $x_{1,2}^s$ is the fraction of strange over non-strange quarks of the two incoming hadrons. The cross section of unknown processes are then scaled with
 \begin{equation}
   \frac{\sigma^{\mathrm{AQM}}_{\mathrm{process}}}{\sigma^{\mathrm{AQM}}_{\mathrm{ref-process}}} \sigma^{\mathrm{AQM}}_{\mathrm{ref-process}} \, .
 \end{equation}
 Here $\sigma^{\mathrm{AQM}}_{\mathrm{ref-process}}$ is a reference process where the actual cross section is experimentally well defined. The inclusion of AQM cross sections will enhance the number of interactions as there are more cross section between hadrons. 

 We additionally incorporate angular distributions on the final state of elastic and inelastic binary interactions of the form
 \begin{equation}\label{Eq:cugnon}
  d\sigma / dt \propto e^{-bt} \, ,
 \end{equation} 
 taken from \cite{Cugnon:1996kh}. Here the baseline is the measurement of the angular distribution in elastic proton proton interaction. Starting with elastic NN processes which distribution is measured, it is argued in \cite{Cugnon:1996kh} that the final state in $NN\rightarrow N\Delta$ scatterings has a similar shape as the elastic NN interaction, which is why we incorporate the same distribution here. Due to the lack of experimental data on the measurements of angular distributions we additionally incorporate the same distributions for all elastic scatterings in our code.

 Introducing these new features into the code we want to examine their influence on the transport coefficients. The ones that are of interest in this work are:
 \begin{itemize}
   \item Multi-particle scatterings according to Eq.~\ref{eq:P31}
   \item Elastic cross sections via the AQM description, see Eq.~\ref{Eq:AQM}
   \item Inclusion of anisotropic scatterings between baryonic scatterings.
 \end{itemize}
 By individually turning these interactions on and off we are able to directly compare their results and their impact on $\eta$ and $\kappa_{ij}$. The two latter ones are used per default whereas the multi-particle scatterings are only included as an exploratory study.
 
\section{\label{methodology}Methodology}
  We extract the transport coefficients shear viscosity $\eta$ and diffusion coefficients $\kappa_{ij}$ by making use of the Green-Kubo formalism which relates the transport coefficient with the auto-correlation function of an equilibrium current for the diffusion coefficient or dissipative flux for the viscosity.
  In the case of the shear viscosity the off-diagonal components of the energy momentum tensor $T^{ij}$ ($i\neq j$, $i,j \in \{x,y,z\}$) are used and the baryon, electric charge and strangeness currents $\vec j_{Q_k}$ with $Q_k\in \{B,Q,S\}$.
  The Green-Kubo relations are then written as
  \begin{align}
    \eta &= \frac{V}{T}\int \langle T^{ij}(t) T^{ij}(0)\rangle dt \label{Eq:GreenKubo_eta} \\
    \kappa_{ij} &= \frac{V}{3} \int \langle \vec J_{i}(t) \vec J_j(0)\rangle dt \, . \label{Eq:GreenKubo_kappa}
  \end{align}
  Here $V$ is the volume of the system and $T$ the corresponding temperature.
  
  For a system based on discrete time steps the auto-correlation can be computed as follows
  \begin{equation}\label{Eq:AutoCorrelation}
    \langle \mathcal{I}(0)\mathcal{I}(t)\rangle = \lim_{K\to\infty}\frac{1}{K-u}\sum_{s=0}^{K-u} \mathcal{I}(s\Delta t)\mathcal{I}(s\Delta t + u\Delta t) \, ,
  \end{equation}
  where $K$ is the total number of time steps, $\Delta t$ the time interval and $u < K$ a positive integer with $t=u\Delta t$. $\mathcal{I}$ is written as a replacement for the charge current $\vec J$ or the shear stress component $T^{ij}$.
  The time evolution of the system is modeled by evolving each single particle distribution function. As a result, the total one can be written as $f(x,p) = \sum_a \delta^{(3)}(x-x_a)\delta^{(3)}(p-p_a)$ with the sum running over all particles in the system.

  As a result the off-diagonal components of the energy momentum tensor are
  \begin{equation}
    T^{ij} = \int \frac{d^3p}{p^0} p^i p^j f(x,p) = \frac{1}{V}\sum_{a=0}^N \frac{p_a^i p_a^j}{p_a^0} \, .
  \end{equation}
  With $p_a^i$ is the $i$-th momentum component of particle $a$.

  Similary, the baryon, electric and strangeness charge currents $\vec j_{Q_k}$ with $Q_k\in \{B,Q,S\}$ are computed as
  \begin{equation}
    j^i_{Q_k} = \int \frac{d^3p}{p^0} Q_k p^i f(x,p) = \frac{1}{V}\sum_a \frac{p_a^i}{p_a^0} Q_{k,a} \, .
  \end{equation}

  The entropy of the system is calculated as
  \begin{equation}\label{Eq:GibbsEntropy}
    s = \frac{w - \sum_{q\in\{B,Q,S\}}\mu_q n_q }{T} \, .
  \end{equation}
  Here $w = \epsilon + P$ is the enthalpy consistent of the energy density $\epsilon$ and pressure $P$. The chemical potential $\mu_q$ and net charge density $n_q$ of the net-baryon $B$, net-charge $Q$ and net-strangeness $S$. The temperature of the system is calculated by comparing the energy density of the most abundant and stable particles to expectations from the equilibrium Boltzmann distribution $f_{\rm{eq}}$.\\\\

  Within this work we calculate the shear viscosity and diffusion coefficients Eq. \ref{Eq:GreenKubo_eta}-\ref{Eq:GreenKubo_kappa} by numerically integrating the auto-correlation function Eq. \ref{Eq:AutoCorrelation}.
  In previous works, the transport coefficients were calculated by fitting the correlation functions by assuming a specific shape (e.g. one or multi-exponential function).
  The method of numerical integration has the benefit that one can drop the assumption of an exponential shape and capture much more information from the correlation functions at larger times. However it has the draw-back that the correlation function must be known very well which can be challenging since their tails suffers from large noise due to lack of statistics.
  Especially at low temperatures ($T<100$ MeV) where the relaxation time of the system is large, the correlation function decreases slowly in comparison to systems at larger temperatures with smaller relaxation times.
  We first compute the correlation functions of each event and then calculate the transport coefficients from the averaged ones.
  We have thoroughly checked the convergence of the integrals depending on the upper limit of time. The error of $\eta$ and $\kappa_{ij}$ are determined by integrating the upper / lower error respectively of the averaged correlation function and are presented as bands in all figures.
  For all calculations, the integral is evaluated until a relative error of $5\%$ is reached.
  
  We take time step size of $\Delta t = 0.05\,\rm{fm}$ and in order to increase the statistics the results of the three spatial components of the charge currents and the three different off diagonal components of the energy momentum tensor are taken as independent events.

  \begin{figure}[h]
    \centering
    \includegraphics[width=0.48\textwidth]{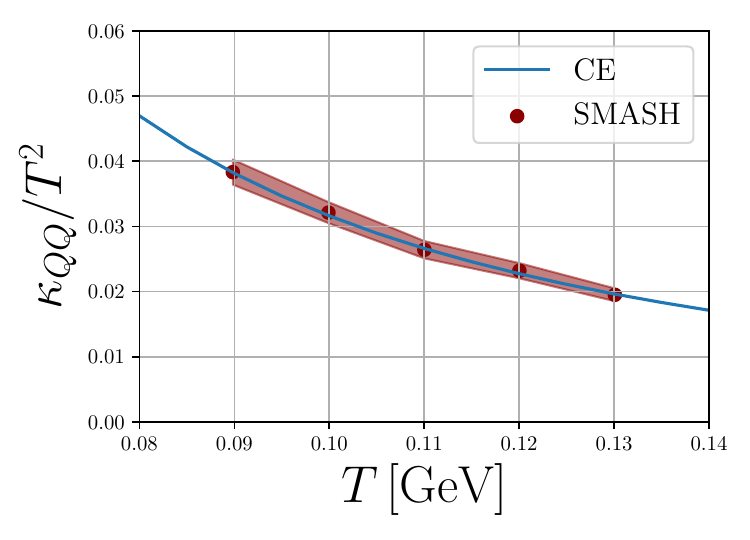}
    \caption{Electric charge diffusion coefficient of a pion gas interacting via a constant isotropic cross section of $30$ mb as function of temperature. The full line shows results from Chapman-Enskog calculation (CE) and the points are results from SMASH.}
    \label{Fig:KappaQQConstPions} 
  \end{figure}
  Fig. \ref{Fig:KappaQQConstPions} shows a test calculation to verify our methodology described above. We run simulations of a pion gas ($\pi^+,\pi^-,\pi^0$) interacting via a constant isotropic cross section of $\sigma_{tot} = 30$ mb and compared the result of Eq. \ref{Eq:GreenKubo_kappa} to Chapman-Enskog calculations \cite{Fotakis:2019nbq}. We find perfect agreement with our refined method of numerical integration of the correlation function.

\section{\label{multipart}Multi-Particle Reactions}
  
  We start with the calculation to quantify the impact of multi-particle reactions on the shear viscosity. For this, a simplified interacting hadron gas is employed consistent of $\pi$, $\rho$ and $\omega$.
  The interaction channel of interest is the multi-particle reaction $3\pi\leftrightarrow\omega$. With the stochastic criterion, it is possible to treat this reaction directly, using the collision probability in Eq.~\ref{eq:P31}. Limiting the calculation to binary scatterings, as required by the geometric collision criterion, the same reaction is described as a reaction chain via an intermediate $\rho$ resoanance: $\omega\leftrightarrow\rho\pi\leftrightarrow 3\pi$. For a fair comparison the $\rho$ meson is added as a degree of freedom to the system independent of the treatment of the $3\pi\leftrightarrow\omega$ reaction. 
  After initializing the system and letting it evolve dynamically with the type of interaction of interest, the correlation functions are calculated and integrated to get the transport coefficient.
  
  \begin{figure}[h]
    \centering
    \includegraphics[width=0.48\textwidth]{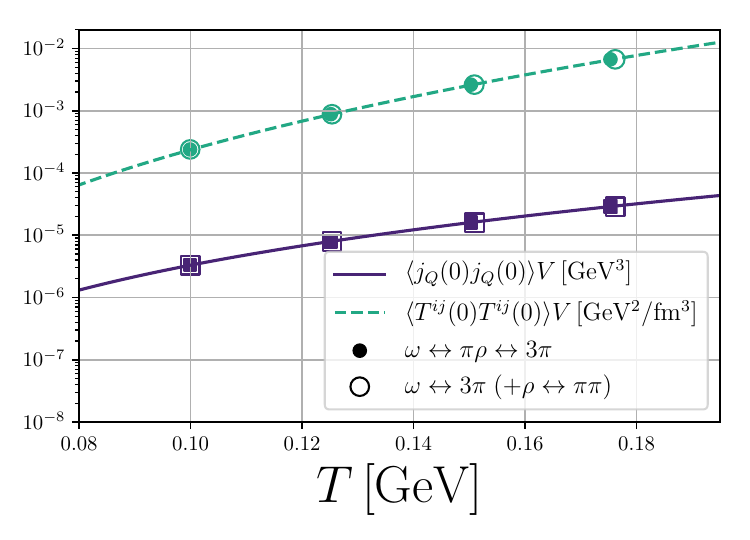}
    \caption{Value of current-current and energy momentum correlation function at $t=0$ as a function of the temperature. Numerical result from SMASH are shown as points and semi-analytical results are presented in lines.}
    \label{Fig:C0MultiPart} 
  \end{figure}
  Fig.~\ref{Fig:C0MultiPart} shows that the value of the correlation function at $t=0$ is equivalent for both cases and matches analytical expectations (see e.g. Eq. 9 in \cite{Rose:2017bjz}).
  As a result, differences in the shear viscosity between the two treatments can only arise from different relaxation times $\tau$.
  \begin{figure}[h]
    \centering
    \includegraphics[width=0.48\textwidth]{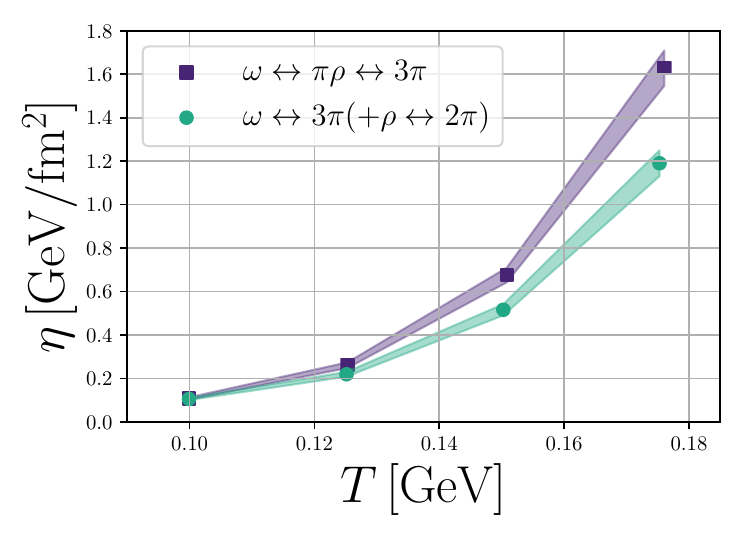}
    \caption{Shear viscosity as a function of the temperature for a simplified interacting hadron gas consistent of $\pi$ $\rho$ $\omega$. The results are shown for the case of multi-particle reactions (green symbols) and geometric collision criterion (purple symbols).}
    \label{Fig:EtaMultiPart} 
  \end{figure}
  \begin{figure}[h]
    \centering
    \includegraphics[width=0.48\textwidth]{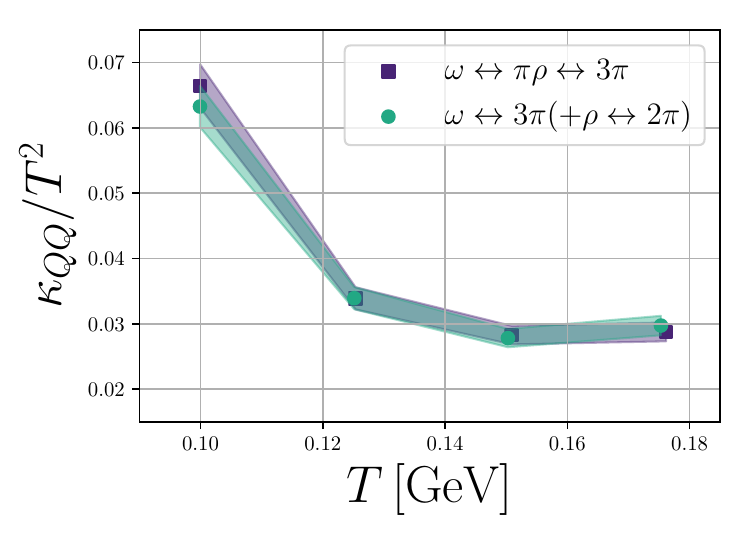}
    \caption{Electric charge charge diffusion coefficient as a function of the temperature of the $\pi$ $\rho$ $\omega$ gas interacting via multi particle reactions (green) and binary reaction chanes (purple).}
    \label{Fig:KappaQQMultPart} 
  \end{figure}
  Fig. \ref{Fig:EtaMultiPart} shows the shear viscosity $\eta$ as a function of the temperature for the two different treatments of interactions. In order to minimize systematic uncertainties both system were calculated using the stochastic collision criterion even though in one of them only the binary interactions were enabled. Both curves show an increase of $\eta$ with increasing temperature, since at larger temperatures the system gets populated by the heavier $\rho$ and $\omega$ mesons. At low temperatures the system is dominated by pions.
  Differences between the two cases become visible above $T\sim 120\,\rm{MeV}$, where the $\omega$ and $\rho$ meson which have similar masses become more important and therefore their interactions play a larger role. At lower temperatures the gas is mostly dominated by pions.
  For the shear-viscosity the main driver is the time with which the off-diagonal components of the energy momentum tensor can relax from small perturbations.
  With the use of multi-particle ($3\leftrightarrow 1$) reactions, the phase space of the outgoing particles opens up much faster and therefore reduces the relaxation time of the off-diagonal components of the energy momentum tensor. In the case of the binary reaction chain, the perturbations need a longer time to equilibrate compared to the $3\leftrightarrow 1$ reactions. The effect of a faster equilibration time with multi-particle reaction was also one of the results of previous studies \cite{Staudenmaier:2020xqr,Staudenmaier:2021lrg}. The differences end up with an increase of the shear viscosity of $\sim 30\%$ at $T=175$ MeV.

  Fig. \ref{Fig:KappaQQMultPart} shows the electric charge diffusion coefficient $\kappa_{QQ} / T^2$. With increasing temperature the diffusion coefficient shows a decreasing trend which can be mainly attributed to the factor $\sim 1/T^2$. The coefficient $\kappa_{QQ}$ itself does not strongly depend on the temperature. 
  Contrary to the viscosity the diffusion coefficient is not affected by the inclusion of the multi-particle reaction.
  Since the value of the correlation function at time zero is equal in both cases, the relaxation time of the current-current correlation function must also be the same for the two systems. In \cite{Fotakis:2019nbq} is was argued that the diffusion coefficients have a strong dependence on the density of the charge carries, the total hadron density and the cross section of the medium. Since non of them is altered by the inclusion of the $3\leftrightarrow 1$ reactions, it is reasonable that there is no difference (the cross section of the $3\leftrightarrow 1$ is chosen to be the same as the cross section of the binary reaction chain).

Even though a comparison for more complex system or even a full hadron gas would be interesting, it is difficult to facilitate a similar direct comparison between multi-particle reactions and reaction chains modeling the same interaction. The main reason is that only a few multi-particle reactions are introduced and in addition, the combinatorics and therefore runtime of including reaction with more particles become a limiting factor for the calculations. 

\section{Result of the full hadron gas}

\subsection{\label{FullSMASH}Shear viscosity}
  Moving on to the shear viscosity of the full SMASH hadron gas.
  Due to improvements e.g. of the description of cross sections in the code and comparisons to experimental measurements of heavy-ion collisions where the knowledge of the shear viscosity is important, we want to give an updated version of $\eta / s$ compared to \cite{Rose:2017bjz}. 
  \begin{figure}[h]
    \centering
    \includegraphics[width=0.48\textwidth]{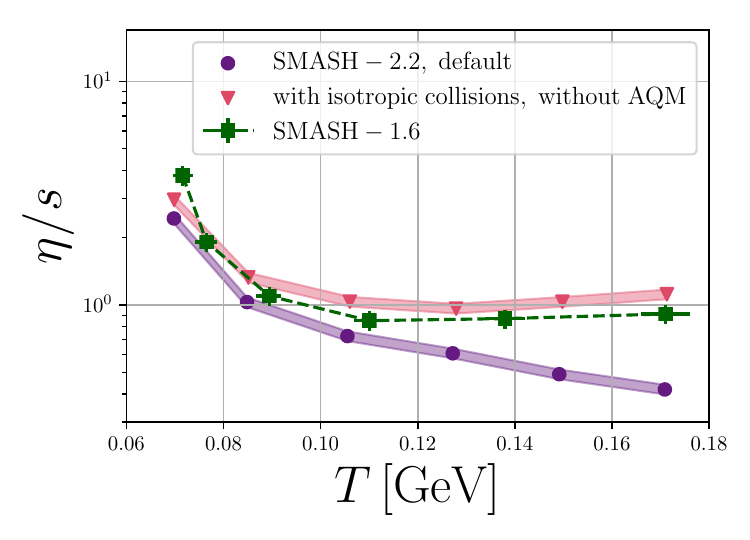}
    \caption{Shear viscosity over entropy density as a function of the temperature of the SMASH hadron gas at vanishing baryon chemical potential. Green squares show the result of the previous version of the code (SMASH-1.6).
    Purple points present the recent value of $\eta / s$ and red triangles the shear viscosity if AQM cross sections and anisotropic scatterings are not used.}
    \label{Fig:EtaOverSFullSMASH}
  \end{figure}

  \begin{figure}[h]
    \centering
    \includegraphics[width=0.48\textwidth]{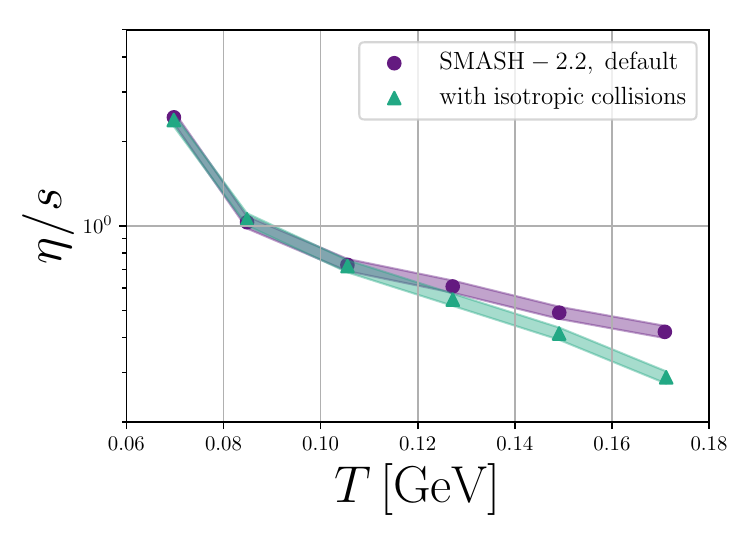}
    \caption{$\eta / s$ as function of the temperature. Results of the SMASH hadron gas are shown as well as the result when all interactions are treated isotropic.}
    \label{Fig:EtaOverSFullSMASHAngular}
  \end{figure}
  Fig. \ref{Fig:EtaOverSFullSMASH} shows the result of the shear viscosity over entropy density of the SMASH hadron gas with the inclusion of improved interaction types in our model as described above. One of the main result is that the inclusion of AQM cross sections drastically decreases the value of $\eta$. The reason being that the estimates of cross sections for hadrons where there is no experimental measurement increases the number of interactions in the hadron gas. This enhanced scattering reduces the relaxation time of the system because there are more channels to equilibrate small pertubations in $T^{ij}$.
  A similar decreasing effect of the shear viscosity with additional interaction channels has been in seen in \cite{Rose:2017bjz} (Fig. 15 top left) where a constant elastic cross section has been added to the system and as a result $\eta / s$ decreased in a similar fashion. 
  The inclusion of angular distribution of the final state in binary interactions does not affect the shear viscosity as strongly as the AQM cross sections. However even though the error bars prevent a clear statement, there seems to be a consistent trend that anisotropic angular distributions increase the shear viscosity, which is consistent with what has been found e.g. in \cite{Lin:2021mdn} where forward scattering in a gluonic medium increase $\eta / s$ in the AMPT model.
  Due to limitations of the phase space distributions of outgoing particles a shear stress cannot relax as fast as in case of isotropic distribution.
  Finally in the case where collisions are treated isotropical and no AQM cross sections are included (red triangles in Fig. \ref{Fig:EtaOverSFullSMASH}) we approximately recover the result from SMASH 1.6 that was published in \cite{Rose:2017bjz}.

\subsection{\label{Diffusion}Diffusion coefficients of conserved charges}
  We now want to discuss the results of baryon, strangeness and electric charge diffusion coefficient matrix. Previously, results of the electric charge diffusion coefficients (QQ, QB, QS) have been published in \cite{Rose:2020sjv}. We thus want to extent this results with the baryonic as well as the strange sector (BB, SS, BS) to complete the picture and investigate the effects of the AQM cross section and anisotropic scatterings.
  \begin{figure}[h]
    \centering
    \includegraphics[width=0.48\textwidth]{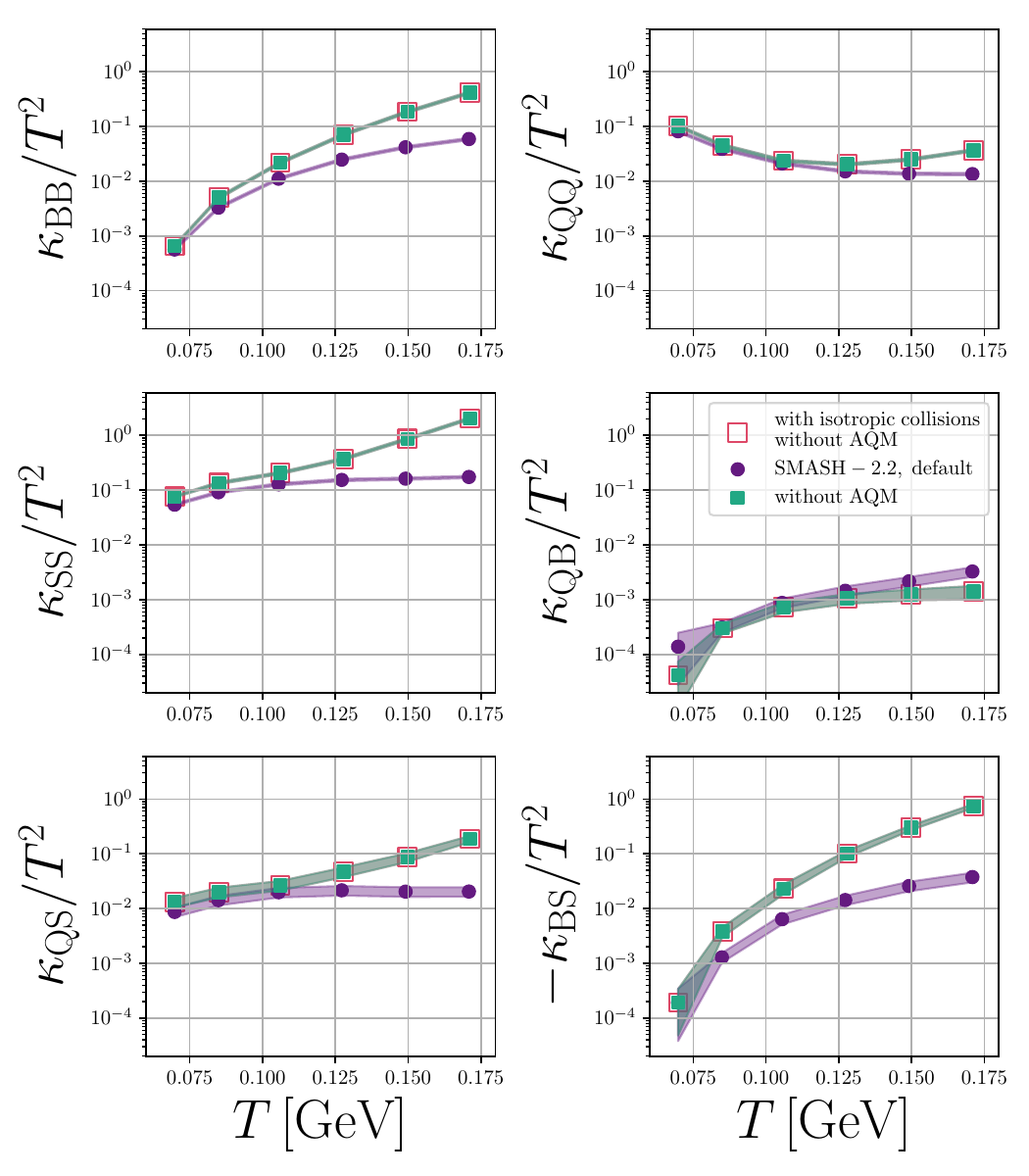}
    \caption{Full diffusion matrix $\kappa_{ij}$ as a function of the temperature at vanishing baryon chemical potential.
    The results of the SMASH hadron gas (purple), the impact of missing AQM cross section (green squares) and missing AQM cross section plus isotropic treatment of collisions (open orange squares) are shown.}
    \label{Fig:DiffMatrixCollTerm}
  \end{figure}

  Fig \ref{Fig:DiffMatrixCollTerm} shows the full diffusion coefficient of conserved charges from SMASH which additionally show the influence of angular distributions and AQM cross sections. 
  For all diffusion coefficients except $\kappa_{QQ}$ we find a monotonic increasing behavior as a function of temperature. This is a result of an increase of the charge density as more charge carriers become available to the system.
  In the case of $\kappa_{BB}$ the lightest charge carrier is the nucleon, for $\kappa_{SS}$ the kaon and in the case of $\kappa_{BS}$ it is the $\Lambda$ baryon which carries both baryon and strangeness charge. However with increasing hadron density the scattering rate increases as well which suppresses the diffusion coefficients.
  
  The additional cross sections from the AQM model have a strong impact on all diffusion coefficients. At low temperatures where the hadron gas is mostly dominated by pions, the inclusion of the AQM cross section play no large role. With increasing temperature however, where the baryonic degrees of freedom start to dominate, the additional elastic cross section significantly decrease the diffusion coefficients.
  This originates from the fact that the diffusion coefficients are proportional to the scattering rate and the cross section $\sim 1 / \Gamma_{scat} \sim 1 / (n_{tot}\sigma_{tot})$ therefore with additional cross sections the coefficients $\kappa_{ij}$ decrease.
  
  Interestingly and in contrast to the shear viscosity $\eta$, non-isotropic distributions have no visible influence on the diffusion coefficient. A possible explanation is the strength of anisotropy is simply not strong enough. Additionally, angular distributions are only present between some baryonic species and no mesonic ones. 

\section{Influence of varying degrees of freedom}\label{Sec:DOF}
  In this section we want to disentangle the effect of having different hadronic degrees of freedom in the system on the transport coefficients $\eta$ and $\kappa_{ij}$. This is particularly interesting since a strong dependence on the degrees of freedom gives rise to a good comparison between this calculation and possible first-principle calculations.
  In the presented calculations we have used three different systems with varying hadron species and interactions and with increasing complexity. A more detailed description is given in Appendix \ref{App:2}.
  We start with a simple version of a hadron gas of stable particles interacting via $2\leftrightarrow 2$ collisions with a constant isotropic cross section of $30$ mb denoted as $\pi K N \Lambda \Sigma$ (const.).
  As a next step we introduce the resonances shown in Tab. \ref{App:TableDoF} and add interaction channels like resonance formation and decay processes $2\leftrightarrow 1$ and inelastic $2\leftrightarrow 2$ processes. The list of these hadrons is similar to the one used in \cite{Fotakis:2019nbq}
  And finally the results of the full hadron gas provided by SMASH that has already been discussed in the previous section. All three different boxes are calculated with $\mu_B = 0$.
  \begin{figure}[h]
    \centering
    \includegraphics[width=0.48\textwidth]{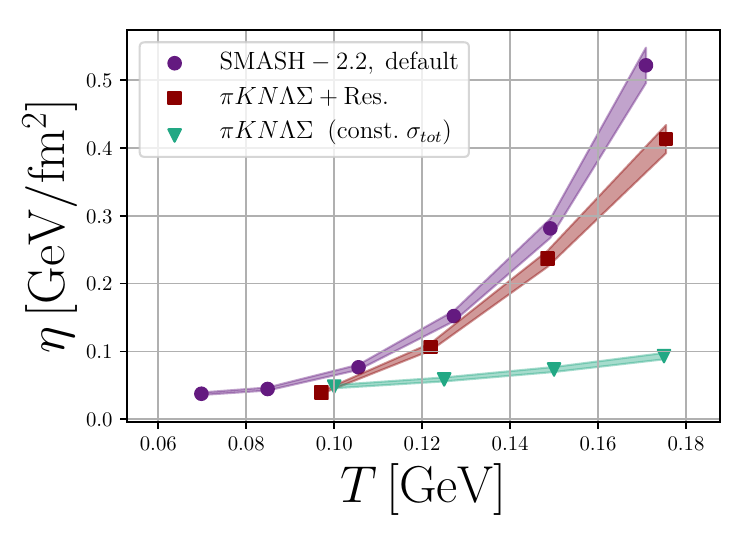}
    \caption{Shear viscosity $\eta$ as a function of the temperature. The results of three different hadronic systems are compared. The hadron gas interacting via elastic cross sections consistent of $\pi K N \Lambda \Sigma$ (green triangles), the hadron gas interacting via more complicated interactions consistent of $\pi K N \Lambda \Sigma$+Res. gas (red squares) and the full SMASH hadron gas (purple points) are presented.}
    \label{Fig:EtaSystems}
  \end{figure}

  \begin{figure}[h]
    \centering
    \includegraphics[width=0.48\textwidth]{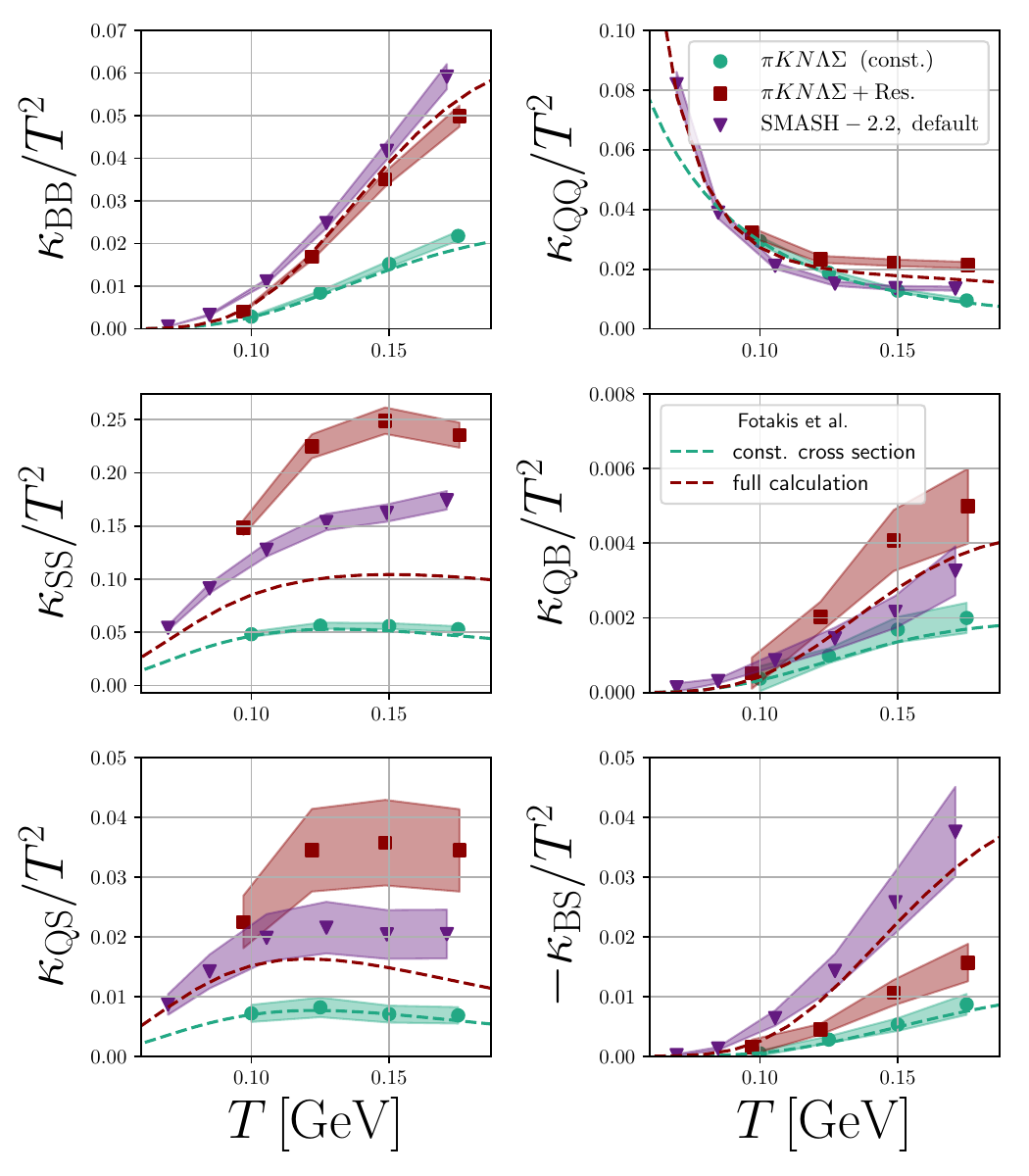}
    \caption{Diffusion coefficients matrix $\kappa_{ij}$ as a function of the temperature. The same hadronic systems as in Fig. \ref{Fig:EtaSystems} are presented.}
    \label{Fig:DiffMatrixBS}
  \end{figure}

  In Fig. \ref{Fig:EtaSystems} the shear viscosity of the three hadronic systems is shown. Whereas the simplest hadron gas $\pi K N \Lambda \Sigma$(const.) has the lowest shear viscosity across the temperature region, the other two calculations SMASH-2.2 and $\pi K N \Lambda \Sigma$+Res. are in similar order of magnitude and their shape. Especially at large temperatures, the difference between the hadron gas interacting via a constant cross section and the other two is significant whereas at lower temperatures the values of $\eta$ seems to converge.
  This result shows that $\eta$ is not strongly dependent on the degrees of freedom but rather on the interactions between them.
  Even though there are much more hadronic species in the full hadron gas than in simplified hadron gas which still includes interactions via resonances, $\eta$ is similar.
  For the gas with a constant cross section there are much more interaction channels at higher temperatures which can relax small pertubations of the off-diagonal components of $T^{\mu\nu}$. In addition for the hadron gases with resonances the non-zero lifetimes of those species become important at large temperatures which effectively increase $\eta$ \cite{Rose:2017bjz}.

  Fig. \ref{Fig:DiffMatrixBS} shows the  diffusion coefficient matrix of the three different hadron gases.
  First, the comparison of $\kappa_{ij}$ of the $\pi K N \Lambda \Sigma$(const.) gas to Chapman-Enskog calculation \cite{Fotakis:2019nbq} matches perfectly. Continuing with the more complicated $\pi K N \Lambda \Sigma$(Res.) gas which consists of the same degrees of freedom as the full result from \cite{Fotakis:2019nbq} we already see a large difference between the two calculations. This already shows that the detailed description of the cross sections has a large influence on $\kappa_{ij}$. E.g. resonances which propagate in space or mass dependent decay widths $\Gamma(m)$ of the unstable particles can play a role. Note that the resonances in the CE calculations are incorporated as Breit-Wigner peaks in the energy dependent cross section.
  It was found in \cite{Rose:2020sjv}, that the three electric charge diffusion coefficients follow the same trend that $\kappa_{QX}$ decreases when the active degrees of freedom were increased. It was argued that with more degrees of freedom the charge density increases and therefore the scattering rate which reduces the diffusion coefficient.
  We find a similar behavior for $\kappa_{SS}$ but not for $\kappa_{BB}$ and $\kappa_{BS}$, where the diffusion coefficients increase when going from the $\pi K N \Lambda \Sigma$+Res. gas to the full SMASH hadron gas throughout the whole temperature region considered.
  In contrast to $\kappa_{QX}$ and $\kappa_{SS}$ the enhancement of the scattering rate does not compensate the increase of the charge density and as a result $\kappa_{BB,BS}$ gets larger.

  \section{Non-zero baryon chemical potential}\label{Sec:FiniteMuB}

    We now want to discuss the behavior of transport coefficients at a non-zero baryon chemical potential $\mu_B$ on both the shear viscosity $\eta$ and the diffusion coefficients $\kappa_{ij}$. For this purpose simulations with $\mu_B = 300$ MeV and $\mu_B = 600$ MeV were performed. In order to mimic the regime of applicability in a heavy-ion collision where the strangeness chemical potential is zero and the isospin chemical potential is approximately zero as well, we initialized the system with a combination of $\mu_{B,Q,S}$ such that we have approximately $\mu_S\approx\mu_{I_3}\approx 0$.
    \begin{figure}[h]
      \centering
      \includegraphics[width=0.48\textwidth]{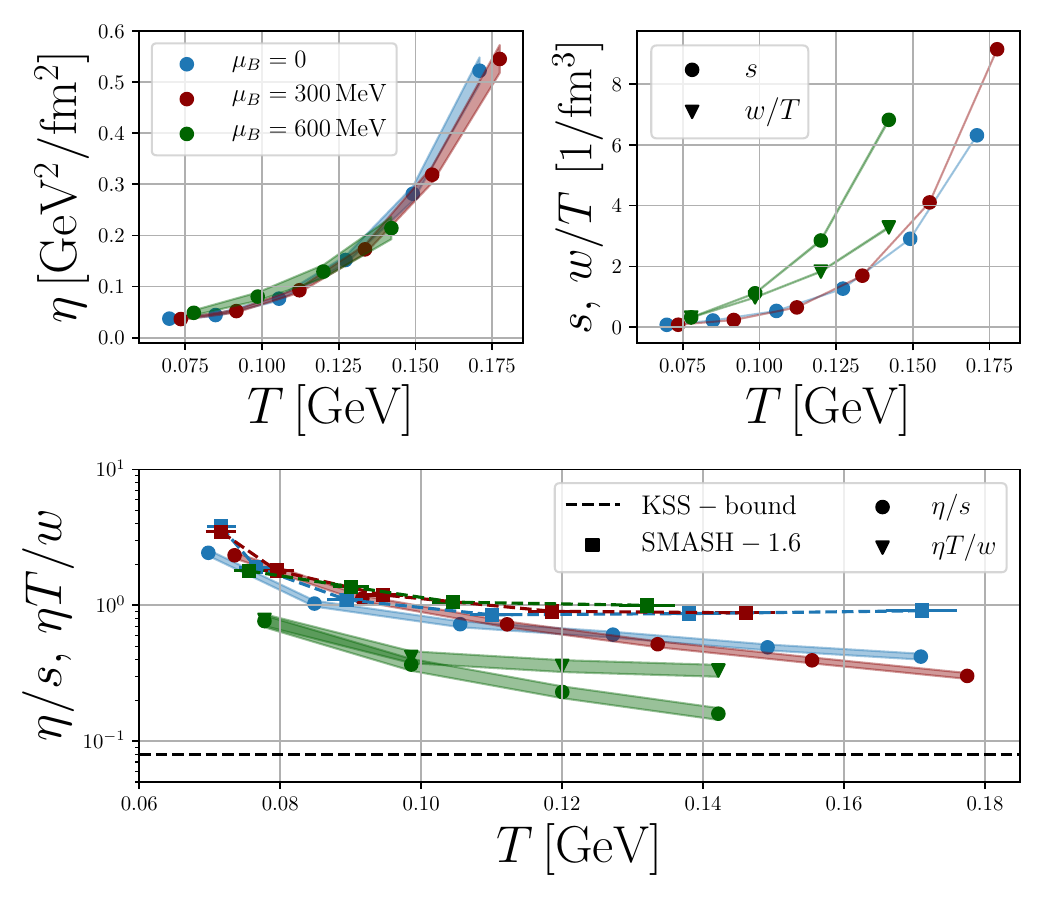}
      \caption{Shear viscosity $\eta$, entropy density (circles) and enthalpy (triangles) and $\eta / s$ (circles) or $\eta T / w$ (triangles) as a function of the temperature. Results are shown for three different values of the chemical potential of $\mu_B = 0$ (blue), $\mu_B = 300$ MeV (red) and $\mu_B = 600$ MeV (green). The results of $\eta / s$ from \cite{Rose:2017bjz} are included as squares and dotted line the KSS bound $1/4\pi$.}
      \label{Fig:EtaOverSFullSMASHMuB}
    \end{figure}

    Fig. \ref{Fig:EtaOverSFullSMASHMuB} shows the entropy density $s$ and enthalpy $w/T$, shear viscosity $\eta$ and $\eta / s$ or $\eta T / w$ for three different values of the baryon chemical potentials. % Additionally the results from SMASH-1.6 are shown.
    The shear viscosity shown in the upper left plot shows no significant dependency on $\mu_B$. In contrast, previous calculation found a strong enhancement of $\eta$ (see Fig. 9 (top) in \cite{Rose:2017bjz}). This change originates from the additional elastic cross sections from the AQM description, which added interaction channels in those baryon rich regimes.
    The dependency of the entropy density $s$ on the chemical potential $\mu_B$ can be seen on the upper right plot in Fig. \ref{Fig:EtaOverSFullSMASHMuB}. We additionally include the enthalpy of the system which is often used instead of the entropy to account for the change in chemical potentials (see Eq. \ref{Eq:GibbsEntropy}). We find that for $\mu_B = 600$ MeV $w / T$ is still enhanced with respect to $s$ at $\mu_B = 0$ and $\mu_B = 300$ MeV.
    The shear viscosity over entropy density or enthalpy has only a weak dependence on the baryon chemical potential up to $\mu_B = 300$ MeV. At the highest value of $\mu_B$ however we observe a strong decrease of $\eta / s$. This completely originates from the enhanced entropy density and not from the transport coefficient itself. In contrast the ratio $\eta T / w$ shows a less strong $\mu_B$ dependence as the chemical potentials are not taken into account.
    Comparing the values of $\eta / s$ to the previous results we find, similar to Fig. \ref{Fig:EtaOverSFullSMASH} a stronger dependence on the baryon chemical potential towards the value of the KSS bound \cite{Kovtun:2004de}.

    \begin{figure}[h]
      \centering
      \includegraphics[width=0.48\textwidth]{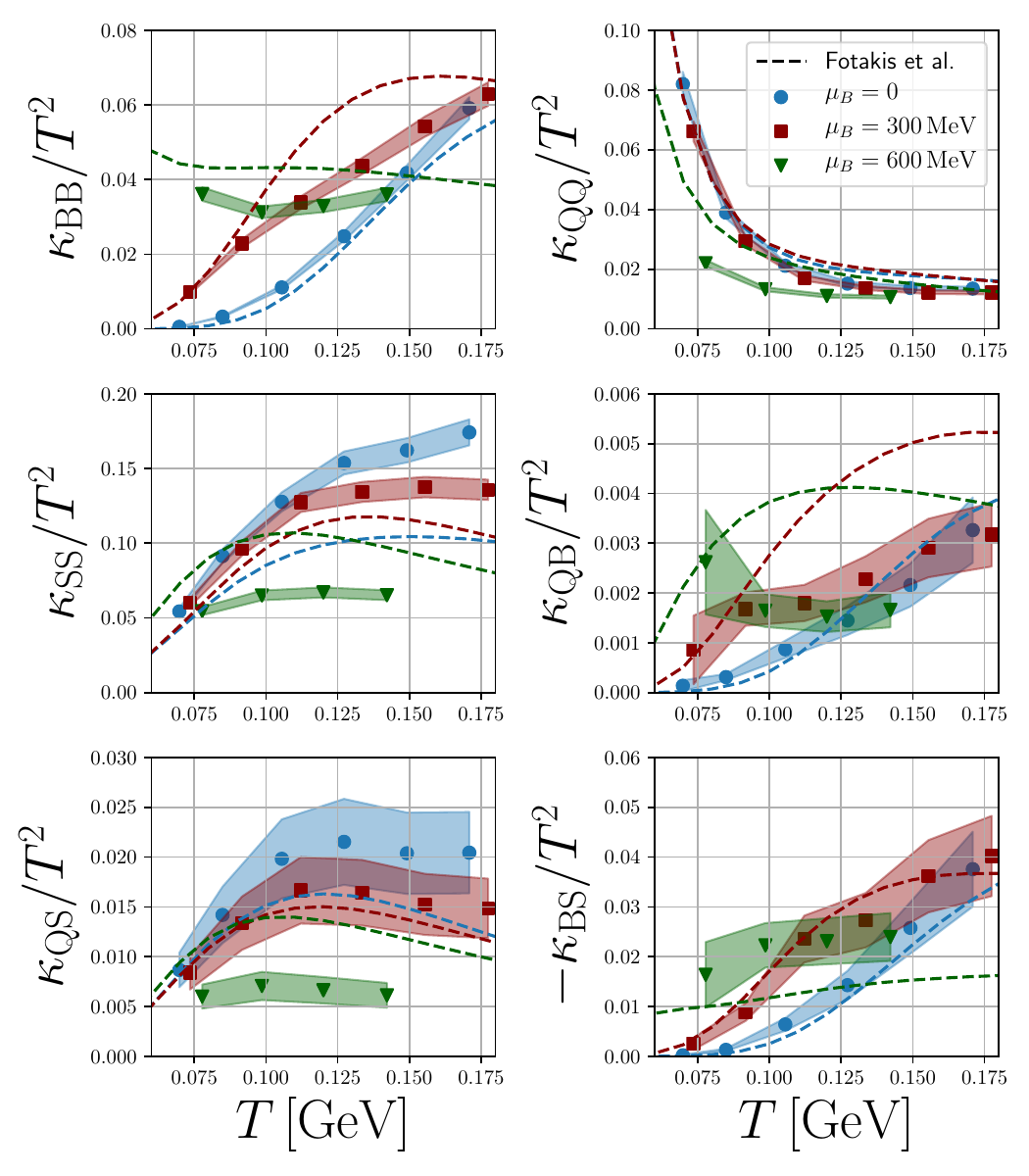}
      \caption{Full diffusion coefficient matrix $\kappa_{ij} / T^2$ as a function of temperature and baryon chemical potential. Results are presented for $\mu_B = 0$ (blue), $\mu_B = 300$ MeV (red) and $\mu_B = 600$ MeV (green). The dashed lines shows results from \cite{Fotakis:2019nbq}.}
      \label{Fig:DiffMatrixMuB}
    \end{figure}
    Now coming to the baryon chemical potential dependence of the full diffusion matrix, which is depicted in Fig. \ref{Fig:DiffMatrixMuB}. In contrast to shear viscosity $\eta$ we find a strong dependency of the whole $\kappa_{ij}$ matrix on $\mu_B$ and we additionally compare our results to the ones from \cite{Fotakis:2019nbq}.
    First of all, the general shape and the orders of magnitude coincide between the two calculations which is sign that they describe similar systems and are robust in terms of their general behavior. When it comes to the exact values of $\kappa_{ij}$ we find that the details of both calculations play a large role. As the degrees of freedom and the description of interactions like resonance formation etc. are very different, it is also expected that the curves don't coincide with each other exactly. This comparison shows that the diffusion coefficients are a good quantity when it comes to comparisons with first principle calculations, but it is also difficult as it depend on counteracting quantities like the relaxation time and (total and net) charge densitites.
    We find no clear trend of $\kappa_{ij}$ when going from 0 to $\mu_B = 300$ MeV. The diffusion coefficients of the baryonic sector $\kappa_{BB}$, $\kappa_{QB}$ and $-\kappa_{BS}$ increase which can also be seen in the CE calculations. The strangeness and electric charge coefficients in contrast show a decreasing behavior.
    We find that for the highest value of $\mu_B = 600$ MeV the diffusion coefficients seem to behave approximately as $\kappa_{ij}\sim T^2$ and as a result the ratio $\kappa_{ij} / T^2$ flattens as a function of temperature. A similar trend can be observed in the CE calculations. 

\section{\label{conclusion}Summary and Conclusions}
  We calculated the influence of multi-particle reactions additional cross sections, number of degrees of freedom and the baryon chemical potential on the transport coefficients $\eta$ and $\kappa_{ij}$. Using a new technique to extract the transport coefficients by integrating the corresponding correlation functions enables us to extract $\eta$ and $\kappa_{ij}$ in a more precise way. We have shown that multi-particle reactions reduce the shear viscosity in a simplified hadron gas at $T\gtrsim 140\,\mathrm{MeV}$ whereas the electric charge diffusion coefficient shows no modification. Continuing with the full SMASH hadron gas it is found that the AQM cross sections strongly decrease both $\eta$ and $\kappa_{ij}$. Additionally the inclusion of angular distributions enhances the shear viscosity but no influence on $\kappa_{ij}$ is found.
  On the other hand, varying the number of degrees of freedom leaves the shear viscosity relatively unaffected until the cross over region $T\approx 150\,\mathrm{MeV}$ in contrast to the diffusion coefficients of conserved charge. Here we find that the scaling behavior that has been observed for the charge sector does not hold in the baryonic sector and both the density and the cross sections become the relevant factors.
  Contrary to previous findings there is no dependency of $\eta$ on $\mu_B$ leaving the dependency of the ratio $\eta / s$ and $\eta T / w$ fully on the thermodynamic properties of the medium. Finally, the $\kappa_{ij}/T^2$ matrix has a strong $\mu_B$ dependence with the tendency of evolving a plateau behavior at $\mu_B\approx 600\,\mathrm{MeV}$.

  For future work, it would be interesting to investigate the behavior of $\eta$ and $\kappa_{ij}$ more in the baryon dense regions. This would include calculations using mean field potentials.

\appendix
\section{Angular distributions}\label{App:1}
  \begin{figure}[h]
    \centering
    \includegraphics[width=0.45\textwidth]{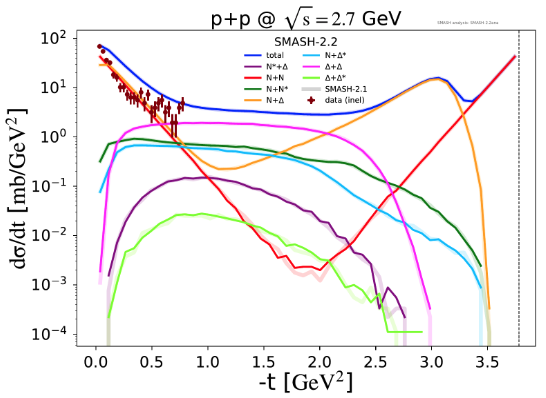}
    \caption{Angular distribution of inelastic pp scatterings. The figure is taken from \cite{SMASH_analysis} and the data points from \cite{Bacon:1967zz}.}
    \label{Fig:AngularDistribution}
  \end{figure}
  Fig.~\ref{Fig:AngularDistribution} shows the angular distribution $d\sigma / dt$ as a function of the Mandelstam variable $t$. The red curve shows the contribution of the elastic $NN$ scattering and it matches well with experimental data. The distribution of heavier resonance states are not symmetric anymore due to restrictions in the phase space distributions. As there are no measurements of the angular distributions of inelastic scatterings we cannot tune the model to them but overall the results are comparable to the measurements.
  The energy dependence of the parameter $b$ in Eq.~\ref{Eq:cugnon} is computed from the parametrization in \cite{Cugnon:1996kh}.\\\\
\section{Particles and its properties}\label{App:2}
  In this section we want to give a more detailed picture of the degrees of freedom which we have used to calculate the transport coefficients.
  \begin{table}[hbt!]
    \begin{tabular}{| c | c | c | c | c | c |}
      \hline
      Particle & Mass $\mathrm{[GeV / c^2]}$ & Degenercy & Width $\mathrm{[GeV]}$ & In system \\\hline\hline
      $\pi$ & $0.138$ & $3$ & 0 & 1,2,3 \\\hline
      $K$ & $0.494$ & $4$ & 0 & 1,2,3 \\\hline
      $\eta$ & $0.548$ & $2$ & $0.548$ & 2,3 \\\hline
      $\rho$ & $0.776$ & $6$ & $0.149$ & 2,3 \\\hline
      $\sigma$ & $0.800$ & $1$ & $0.400$ & 2,3 \\\hline
      $K^\star(892)$ & $0.892$ & $8$ & $0.050$ & 2,3 \\\hline
      $N$ & $0.938$ & $8$ & 0 & 1,2,3 \\\hline
      $\Lambda$ & $1.116$ & $2$ & 0 & 1,2,3 \\\hline
      $\Sigma$ & $1.189$ & $12$ & 0 & 1,2,3 \\\hline
      $\Delta$ & $1.232$ & $32$ & $0.117$ & 2,3 \\\hline
      $\Sigma(1385)$ & $1.385$ & $24$ & $0.036$ & 2,3 \\\hline
      $N(1440)$ & $1.440$ & $8$ & $0.350$ & 2,3 \\\hline
      $N(1520)$ & $1.520$ & $8$ & $0.110$ & 2,3 \\\hline
      $N(1535)$ & $1.535$ & $8$ & $0.150$ & 2,3 \\\hline
      $N(1650)$ & $1.650$ & $8$ & $0.125$ & 2,3 \\\hline
      See \cite{dmytro_oliinychenko_2020_4336358} & & & & 3 \\\hline
    \end{tabular}
    \caption{Hadronic degrees of freedom used in the presented calculations. The degeneracy is the product of spin, charged and anti-particle states.}
    \label{App:TableDoF}
  \end{table}
  Tab. \ref{App:TableDoF} shows the different particles with their most important properties. We have used three different systems in order to explain the influence of various degrees of freedom and their respective interaction on the transport coefficient. System 1 is the simplest system consistent only of the stable particles of Tab. \ref{App:TableDoF} interacting via a constant elastic cross section. This system is also denoted as $\pi K N\Lambda\Sigma$(const.).
  System 2 consists of the table shown in this section and it contains more resonances. In contrast to system 1 one can say that the elastic interactions of the stable particles are exchanged with the formation of various different resonances.
  System 3 is the gas consistent of all hadronic degrees of freedom implemented in the transport code used in this work. The interested reader is referred to \cite{dmytro_oliinychenko_2020_4336358}.

\begin{acknowledgments}
  The authors would like to thank Jan Fotakis and Juan Torres Rincon for fruitful discussions and Jan Fotakis for providing the results of the Chapman-Enskog calculation.

  This work was supported by the DFG SinoGerman project - Project number 410922684. Computational resources have been provided by the Center for Scientific Computing (CSC) at the Goethe-University of Frankfurt. H.E. acknowledges the support by the State of Hesse within the Research Cluster ELEMENTS (Project ID 500/10.006).
\end{acknowledgments}

\bibliography{bibliography.bib}

\end{document}